\documentclass[aps,prl,reprint,superscriptaddress,showpacs]{revtex4-1} 

\usepackage[pdftex]{graphicx}
\usepackage{amsmath}
\usepackage{color} 

\begin{document}

\title{Entanglement of two superconducting qubits in a waveguide cavity via monochromatic two-photon excitation}

\author{S. Poletto}
\author{Jay M. Gambetta}
\author{Seth T. Merkel}
\author{John A. Smolin}
\author{Jerry M. Chow}
\author{A.D. C\'orcoles}
\author{George A. Keefe}
\author{Mary B. Rothwell}
\author{J.R. Rozen}
\author{D.W. Abraham}
\author{Chad Rigetti}
\author{M. Steffen}
\affiliation{IBM T.J. Watson Research Center, Yorktown Heights, NY 10598, USA}

\date{\today}

\begin{abstract}
We report a system where fixed interactions between non-computational levels
make bright the otherwise forbidden two-photon $|00\rangle \rightarrow |11\rangle$
transition. The system is formed by hand selection and assembly of two
discrete component transmon-style superconducting qubits inside a rectangular
microwave cavity. The application of a monochromatic drive tuned to this transition induces two-photon Rabi-like oscillations between the ground and doubly-excited states via the Bell basis. The system therefore allows all-microwave two-qubit universal control with the same techniques and hardware required for single qubit control. We report Ramsey-like and spin echo sequences with the generated Bell states, and measure a two-qubit gate fidelity of $F_g=90\%$ (unconstrained) and $86\%$ (maximum likelihood estimator).
\end{abstract}

\pacs{03.67.Ac, 03.67.Bg, 42.50.Pq, 85.25.-j}

\maketitle

Recent improvements in the qubit coherence times \cite{Paik11,Rigetti12} and fidelities of one- and two-qubit gates \cite{Chow12,Magesan12} for superconducting circuits have fed optimism for large scale quantum information processing with these devices. One-qubit gates are typically performed with exclusively microwave control pulses \cite{Martinis02,Wallraff2005}. These techniques were recently extended to two-qubit gates with the cross-resonance scheme \cite{Paraoanu06,Rigetti10, Mooij10,Chow11,Chow12}. Earlier work required dc tuning of qubit frequencies \cite{Steffen2006,Majer2007,Martinis10,Dewes12}. In particular, references \cite{Strauch03,DiCarlo09, Yamamoto2010,DiCarlo2010} exploited interactions of higher levels of the quantum circuits to produce an effective interaction in the computational basis; the physics of higher-levels has also been exploited elsewhere in quantum computing \cite{Lanyon2009}. Superconducting circuits can be designed to have particular values of their energy transitions and associated derivatives. In this letter we use this capability---in conjunction with the modularity of individual discrete devices within a three-dimensional enclosure---to implement a new all-microwave two-qubit gate induced by the direct drive of the $|00\rangle \rightarrow |11\rangle$ transition, which would be forbidden were it not for the interaction of higher levels.

This transition is impossible in  harmonic systems and is a
small third order interaction  in coupled qubit systems.
However, as we show, it can be made bright in coupled multilevel
systems when the qubit-qubit detuning approaches the anharmonicity.
This transition has also been observed spectroscopically for a two-level
system coupled to a phase qubit \cite{Bushev2010}.
A microwave pulse tuned to this two-photon transition induces an
effective Hamiltonian which implements a rotation in the
$\{|00\rangle,~|11\rangle\}$ subspace whose angle is set by the
action of the pulse, allowing the direct generation of entanglement
from the ground state. The gate is similar to that proposed by
M\o{}lmer and Sorenson \cite{Sorensen2000} which is a bichromatic
two-photon excitation and has become commonplace in trapped ion quantum
computing \cite{Benhelm2008,Monz2011}. As such, it holds promise for the direct generation of entangled states of larger multi-qubit systems.

\begin{figure}[b]
\includegraphics{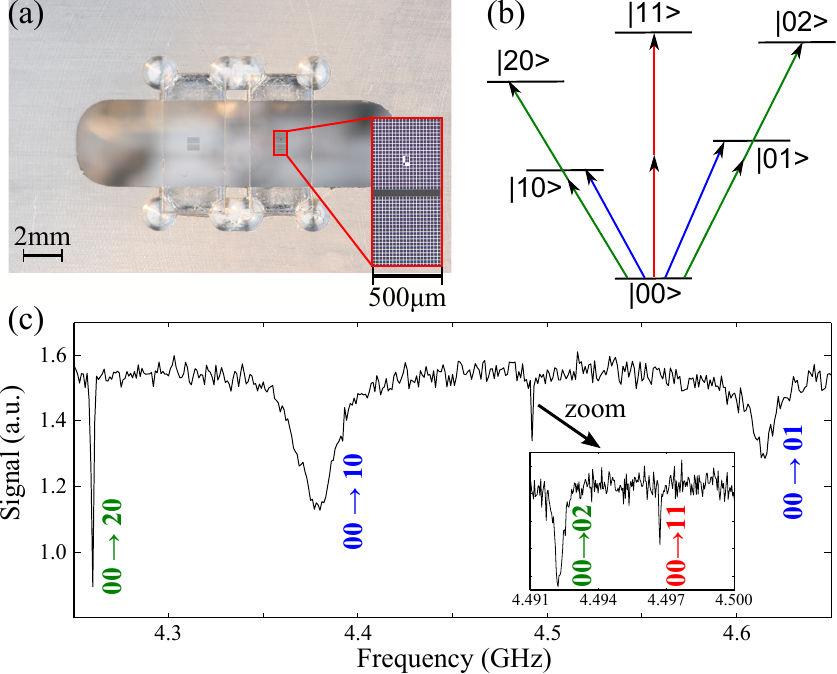}
\caption{\label{fig1} (color online). (a) Picture of half a cavity
with two independent transmons mounted. The inset is a magnified image
of one of the transmons.
(b) Energy diagram of the system. Single and two-photon transitions are depicted
respectively with one and two arrows.
(c) Spectrum of the system. The activation of the transition
$|00\rangle \rightarrow |11\rangle$ requires more power than any other transitions
here measured because it is a second order transition further reduced by a factor of $J/\Delta$.
The dressed cavity resonance frequency is at $11.7781$~GHz. 
}
\end{figure}

The device we study is based on a three-dimensional circuit QED architecture \cite{Paik11}. We leverage the modularity of this design to build up a multi-qubit system from individual discrete components, each of which is independently designed, tested, characterized---and selected---for optimal parameters to realize this effect. 

We use this procedure to implement a two-transmon circuit-QED system \cite{Koch2007,Blais07} described by the Hamiltonian ($\hbar = 1$)
\begin{equation}
\begin{split}
H =& (\omega_1 - \frac{\delta_1}{2}) a^\dagger a + \frac{\delta_1}{2} (a^\dagger a)^2 +(\omega_2 - \frac{\delta_2}{2}) b^\dagger b + \frac{\delta_2}{2} (b^\dagger b)^2\\&+J(a^\dagger b + a b^\dagger)+ \Omega\cos(\omega_d t + \phi)(a+a^\dagger+\lambda b+\lambda b^\dagger)\label{eq:hamiltonian}
\end{split}
\end{equation}
where $\omega_{1(2)}$  is the $|0\rangle \rightarrow |1\rangle$ transition frequency of transmon 1 (2);
$\delta_{1(2)}$ is the anharmonicity of transmon 1 (2);
$J$ is the effective strength of the exchange interaction between transmons;
$\Omega$ is the amplitude of the applied microwave field of frequency $\omega_d$ and phase $\phi$; $\lambda$ is the coupling coefficient of the driving signal
to transmon 2 normalized to transmon 1 (in our case $\lambda\simeq 1$);
and $a$ $(b)$ is the annihilation operators for transmon 1 (2).

In a frame rotating with a drive of frequency $\omega_d $ near the midpoint of
$\omega_1$ and $\omega_2$ the states $|00\rangle$ and $|11\rangle$
form a low energy manifold.
As shown in the supplementary material we can use a sequence of Schrieffer-Wolff
transformations which remove the coupling between this lower energy manifold
and the remainder of the Hilbert space.
Doing so results in an effective Hamiltonian that both couples $|00\rangle \rightarrow |11\rangle$
and produces a Stark shift of each of the four computational states.
By adjusting $\omega_d$ it is possible to make the Stark shifted $|00\rangle $ and $|11\rangle$
levels degenerate in the rotating frame, and thereby generate an effective unitary
$U = U_{B} U_{ZZ}U_{IZ-ZI}$ where
\begin{equation}\label{eqn.bellfreq}
U_B(t) =  \begin{pmatrix}
\cos(\frac{\Omega_{B} t}{2})  & 0 & 0 & -i e^{-2 i \phi} \sin(\frac{\Omega_{B} t}{2}) \\
0 & 1 & 0 & 0 \\
0 & 0 & 1 & 0 \\
-i e^{2 i \phi}\sin(\frac{\Omega_B t}{2})  & 0 & 0 &  \cos(\frac{\Omega_{B} t}{2}) \\
\end{pmatrix}
\end{equation} with
\begin{equation} \label{eqn.bellfreq2}
\Omega_B=\frac{-2J\Omega^2(-J\lambda(\delta_1+\delta_2) +\lambda^2\delta_2(\delta_1+\Delta)+ \delta_1(\delta_2-\Delta)) }{ (\delta_2-\Delta)(\delta_1+\Delta)\Delta^2}
\end{equation}
and $\Delta =\omega_1 - \omega_2$. The operator $U_B$ generates a Rabi-like rotation
at angular frequency $\Omega_B$ about an axis defined by the azimuthal angle $\varphi = 2\phi +\pi/2$ in the equatorial plane of a Bloch sphere whose poles are $|00\rangle$ and $|11\rangle$.
The remaining transformations $U_{ZZ}= \exp(-i \alpha_{zz} ZZ t/4)$
and $U_{IZ-ZI}=\exp(-i \alpha_{-} (IZ-ZI) t/4)$ commute with $U_B$ and, despite the rather complicated
equations for $\alpha_{zz}$ and $\alpha_{-}$, can be left out and corrected with
post processing or refocusing techniques. 

A gate that is locally equivalent to $\mathsf{iSWAP}$ is implemented by choosing
a time $t = \pi/\Omega_B$. We refer to this gate as the $\mathsf{bSWAP}$.
This Clifford gate,
along with single qubit unitaries, forms a universal set of gates for fault-tolerant
computation. With a time $t = \pi/2\Omega_B$ it implements instead a
$\pi/2$ rotation (the $\sqrt{\mathsf{bSWAP}}$ gate), which when applied to
the ground state produces the Bell state $\frac{1}{\sqrt{2}}\left(|00\rangle + e^ {i\varphi} |11\rangle\right)$.
This gate, accordingly, is locally equivalent to $\sqrt\mathsf{iSWAP}$.

\begin{figure}[t]
\includegraphics{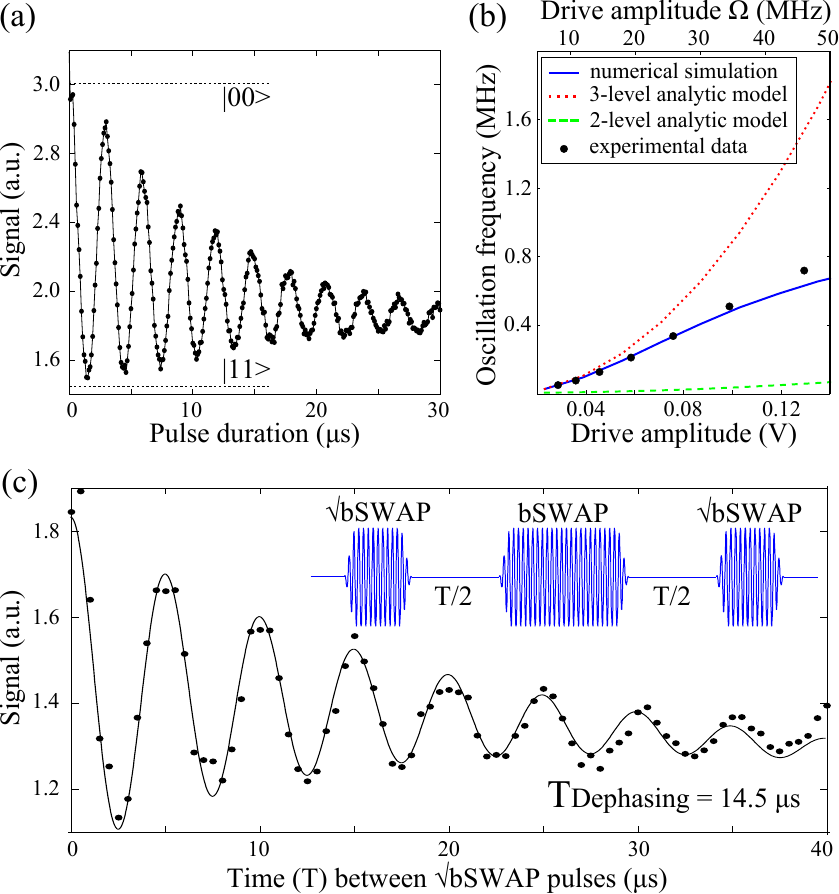}
\caption{\label{fig2} (color online) (a) Two-photon coherent oscillation between the states
$|00\rangle$ and $|11\rangle$. The oscillation
saturates at approximately $1/4$ of its visibility because in the
steady state all four states $|00\rangle$ through $|11\rangle$
are equally populated.
(b) Oscillation frequency versus driving amplitude. Black dots are
experimental data; solid blue line is a multilevel numerical simulation; dotted
red line is derived from perturbation theory of coupled multilevel systems;
and dashed green line is from perturbation theory applied to
coupled two-level systems. We ascribe the discrepancy observed at
large amplitudes to the higher levels of the system, an effect that
is captured by the numerical simulation.
(c) Evaluation of the dephasing time of the Bell state with a spin-echo experiment.
The refocusing $\mathsf{bSWAP}$ is symmetrically placed between
two $\sqrt{\mathsf{bSWAP}}$ gates (inset), and the
oscillation is experimentally induced by linearly ramping the phase of the last pulse in time. 
}
\end{figure}

The magnitude of $\Omega_B$ has some interesting limits.  In the limit of $\delta_i \ll \Delta$ (the harmonic oscillator limit) $ \Omega_B\rightarrow 0$ as expected, while for $\delta_i \gg \Delta$ (pure qubit limit) it reduces to $\Omega_B = -2J\Omega^2(1+\lambda)/\Delta^2$ and for typical values of $\Omega$, $J$, and $\Delta$ it is extremely small, explaining why this effect has not previously been exploited.
The relevant case for this work is the limit in which $\Delta$ approaches
either -$\delta_1$ or $\delta_2$ since the rate passes
through a resonance and becomes large. Such condition is met when the
$\vert 0 \rangle \rightarrow \vert 1 \rangle$ transition of one transmon
approaches the $\vert 1 \rangle \rightarrow \vert 2 \rangle$ transition of the other.
As a consequence the energy level $\vert 11\rangle$ is close to $\vert 02\rangle$
and the leakages from the computational subspace are increased.
It is essentially this leakage rate that determines the maximum
$\Omega_{B}$ with which the gate can be operated.
In this work $\delta_2$ is close to $\Delta$ which results in an enhancement of
$\Omega_B$ by a multiplicative factor of $\delta_2/2(\delta_2-\Delta)$
relative to the pure qubit limit \cite{Note1}.

The enclosure, machined from bulk oxygen-free high thermal conductivity copper
subsequently sputtered with aluminum, has an interior volume of
$(15.5\times 18.6\times 4.2)~$mm$^3$.
Two commercial bulkhead
SMA connectors provide an input and output port to introduce drive signals and perform measurements.
The transmons and the method of applying control pulses through the
enclosure are described in \cite{Paik11,Rigetti12}.
Transmons are fabricated on individual sapphire chips and, following independent pre-characterization,
are selected to match the resonance conditions described earlier (Table \ref{tab:valuesQubits}).
The two chips are mounted symmetrically into the cavity, each 2.1~mm away from the maximum of the $TE_{101}$ mode (see Fig. \ref{fig1}(a)).

\begin{table}
\caption{\label{tab:valuesQubits}
Measured properties of the two transmons independently measured (lines 1 and 2), as well as when placed inside the enclosure (lines 3 and 4).}
\begin{ruledtabular}
\begin{tabular}{ccccccc}
Transmon & Size pads    & $E_{01}$ & $E_{12}$ & $T_1$    & $T_2^*$    & $E_j/E_c$ \\
         & ($\mu$m$^2$) & (GHz)    & (GHz)    & ($\mu$s) & ($\mu$s) &           \\
\hline
A5 (ind) & $500\times 500$ & 4.4513 & 4.2130 & 26 & 16 & 48 \\
A8 (ind) & $600\times 300$ & 4.7013 & 4.4616 & 26 & 13 & 53 \\
A5 (coup) & $500\times 500$ & 4.3796 & 4.1403 & 38 & 29.5\footnote{we report here the $T_2$ value obtained with a spin echo procedure since we observed beating on Ramsey oscillations} & 47 \\
A8 (coup) & $600\times 300$ & 4.61368 & 4.3709 & 32 & 16 & 50
\end{tabular}
\end{ruledtabular}
\end{table}

The sample is shielded by a Cryoperm can filled with Eccosorb$^\textrm{\textregistered}$
foam to suppress spurious thermal radiation \cite{Corcoles2011}.
The signal transmitted by the cavity passes through two circulators and a
low-pass filter at base temperature prior to being amplified by a low noise 
cryogenic HEMT amplifier at 2.8 K and further amplified at room temperature.
The information about the state of the system is extracted by a joint readout
technique \cite{Filipp09} with heterodyne detection by down-conversion.

A spectroscopic measurement of the coupled system is shown in Fig.~\ref{fig1}(c). The strength of the two-photon transition associated with $U_B$ is increased by a factor of 15 compared to the pure qubit limit, greatly enhancing its visibility.
The $ZZ$ coupling is $E_{11}-E_{10}-E_{01}\simeq 90$~kHz and is small enough to
allow for high fidelity single qubit gates with minimal pulse shaping.

Driving the $\vert 00 \rangle \rightarrow \vert 11 \rangle$ transition
generates coherent oscillations between the two states (see Fig. \ref{fig2}(a)). The measured oscillation frequency $\Omega_\mathrm{B}$ versus the amplitude of the microwave driving signal (Fig.~\ref{fig2}(b)) has the quadratic dependence expected
from Eq.~\eqref{eqn.bellfreq2} for small amplitudes. We ascribe the discrepancy observed at higher amplitudes to the higher
levels of the system. To confirm this, we did numerical simulations including the first 3 levels; the results agree quantitatively with our experimental data.

\begin{figure}[b]
\includegraphics{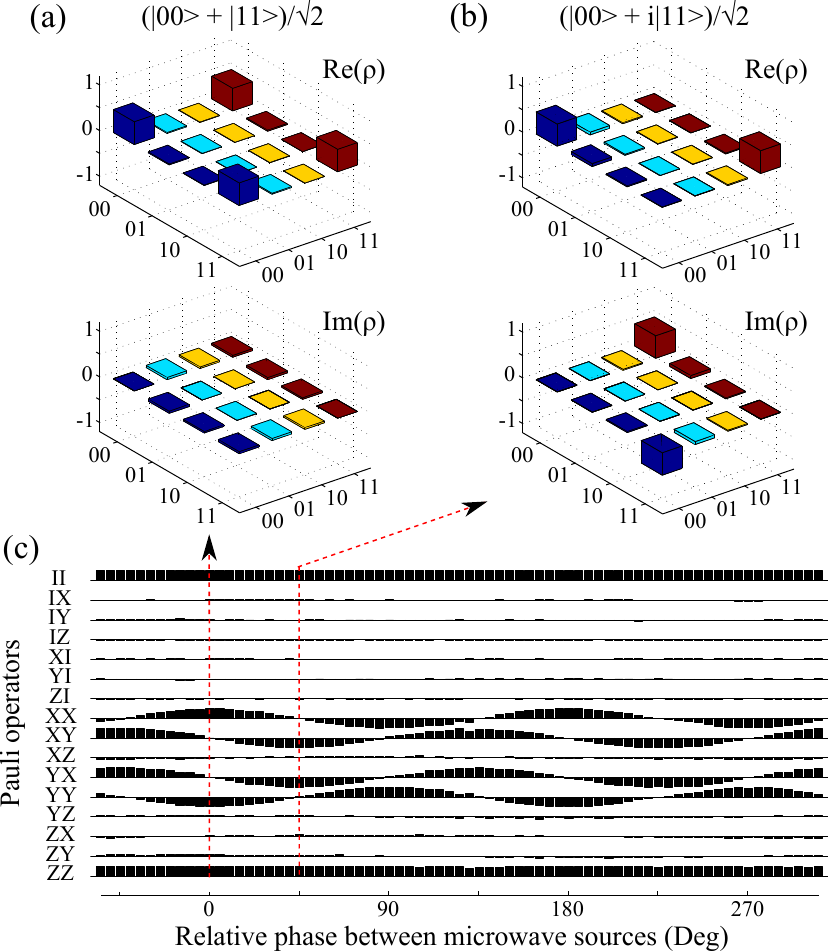}
\caption{\label{fig3} (color online)(a-b) Tomographic reconstruction of the two
Bell states $\frac{1}{\sqrt{2}}\left(|00\rangle + |11\rangle\right)$ and
$\frac{1}{\sqrt{2}}\left(|00\rangle + i|11\rangle\right)$.
(c) Expectation value of the two-qubit Pauli operators versus the relative
phase between microwave sources used to generate the $\sqrt{\mathsf{bSWAP}}$ gate
and the single qubit pulses.
}
\end{figure}

Full state and process tomography on this system requires three microwave
tones to generate the two single qubit gates and the $U_B$ transformation.
Each tone has a phase, and it is necessary to maintain the phase relationship
between $\varphi$ (the phase defining $U_B$) and the single qubit phases.
We approach this problem by using only two microwave sources and single sideband modulation. One microwave source is at frequency $\nu_1=4.49368$ GHz and generates the single qubit pulses employing sideband modulation at -114~MHz and 120~MHz to match the qubit frequencies.
To prevent leakage from the computational subspace the pulses have
a duration of 200~ns and a Gaussian shape with standard deviation $\sigma = 50$~ns.
The other microwave source is at frequency $\nu_2=4.59698$ GHz and generates $U_B$, or the $\sqrt{\mathsf{bSWAP}}$ gate, also via sideband modulation at -100~MHz. To ensure proper phase relationship between individual experiments making up an ensemble average, we obey the experimental repetition rate of $m/(2(\nu_1 - \nu_2))$ where $m$ is an integer.

A Bell state of the form $\frac{1}{\sqrt{2}}\left(|00\rangle + e^{i\varphi} |11\rangle\right)$
is produced by stopping the driven evolution under $U_B$ at the time
$t = \pi/2\Omega_B$ which in this experiment is 800~ns.
Tomographic reconstruction of the prepared state is obtained
by measuring all 36 combinations of the single qubit rotations
$\{I,X_{\pi},X_{\pm\pi/2},Y_{\pm\pi/2}\}^{\otimes 2}$
(with $X_{\alpha}\left[Y_{\alpha}\right]$ being $X \left[Y\right]$ rotations
of $\alpha$ radians). Then either a linear inversion (by a pseudo inverse)
or a maximum likelihood estimation (MLE) \cite{Paris04,Chow12} is used to reconstruct the quantum state. 

We used this procedure to create and characterize the Bell states
$\frac{1}{\sqrt{2}}\left(|00\rangle + |11\rangle\right)$ and
$\frac{1}{\sqrt{2}}\left(|00\rangle + i|11\rangle\right)$ (Fig.~\ref{fig3}(a) and \ref{fig3}(b)).
Measured values for the state fidelities are $>99\%$.

The angle $\varphi$
is completely defined by the relative phase $\phi$ between the two
microwave sources. In Fig. \ref{fig3}(c) the expectation value of the two-qubit Pauli operators
are plotted versus the relative phase between microwave sources. Only the operators
$XX, XY, YX, YY$ have oscillating expectation values,
clearly indicating a transfer of information between real and imaginary
part. The periodicity is double the relative phase between the two
microwave sources as expected from Eq. \eqref{eqn.bellfreq}.

\begin{figure}[t]
\includegraphics{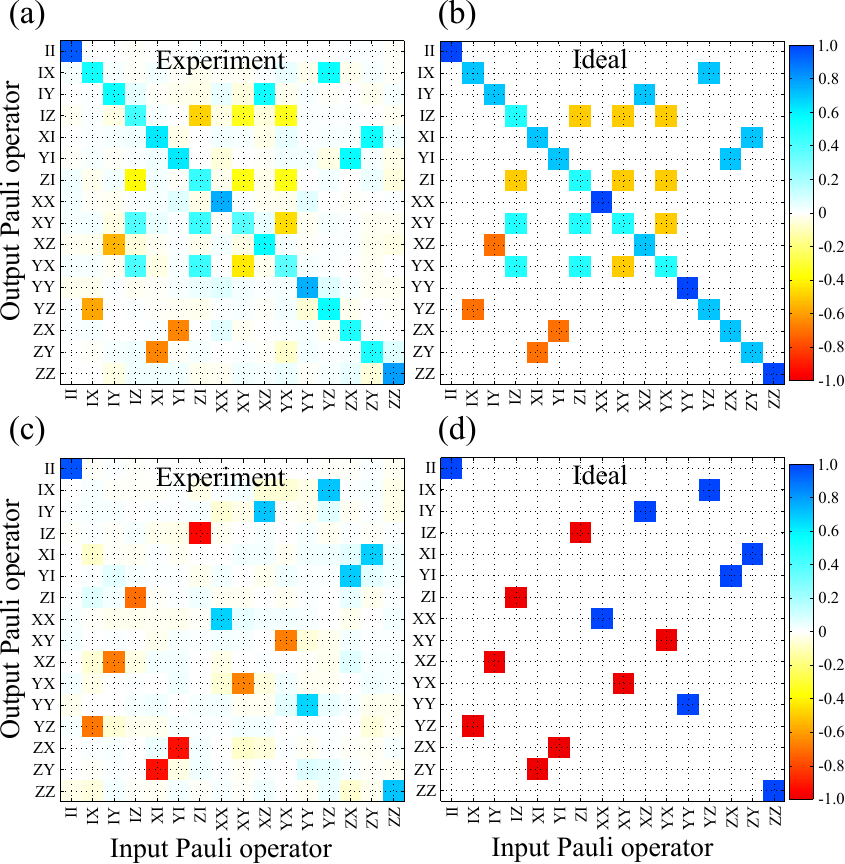}
\caption{\label{fig4}
(color online) Experimental [(a) and (c)] and ideal [(b) and (d)]
Pauli transfer matrices for the $\sqrt{\mathsf{bSWAP}}$ [(a) and (b)]
and the $\mathsf{bSWAP}$ [(c) and (d)] gates.
}
\end{figure}

The ability of the $\sqrt{\mathsf{bSWAP}}$ gate to generate a Bell state
with a single pulse allows for a measurement of the Bell state dephasing
time with a spin echo technique. By implementing the pulse sequence
in Fig.~\ref{fig2}(c)-inset, we measured a Bell state dephasing time of
$14.5~\mu$s (Fig.~\ref{fig2}(c)).

Quantum process tomography of the $\sqrt{\mathsf{bSWAP}}$ and $\mathsf{bSWAP}$ gates is done
by performing a set of $36\times 36$ measurements corresponding to
full state tomography on the 36 different input states generated by
$\{I,X_{\pi},X_{\pm\pi/2},Y_{\pm\pi/2}\}^{\otimes 2}$.
The Pauli transfer matrix $\mathcal{R}$ \cite{Chow12}
is calculated by either a linear inversion or a MLE, and the gate fidelity
is evaluated from the equation
$F_g = \left(\textrm{Tr} \left[\mathcal{R}^{\dagger}\mathcal{R}\right]+4\right)/20$ \cite{Chow12}.
Fig. \ref{fig4}(a) and \ref{fig4}(b) are, respectively, the measured and
ideal Pauli transfer matrices $\mathcal{R}$ for the $\sqrt{\mathsf{bSWAP}}$ gate.
The measured gate fidelity is $F_{\sqrt{\mathsf{bSWAP}}} =90\%$ (raw) and $86\%$ (MLE).

The main source of gate error is a reduction in the dephasing time $T_2^*$ during the experiment induced by the comparatively high power pulse used for the $\sqrt{\mathsf{bSWAP}}$ gate.
This is fit very well by simulations of the Pauli transfer matrix with a $T_2^*$ of $4~\mu$s, but more importantly can be observed directly in the experiment as follows.
We measure $T_\mathrm{echo}$ of qubit A5 in a spin-echo experiment where during the delays we drive the system with a pulse far detuned from any transition but of the same amplitude as that used for the $\sqrt{\mathsf{bSWAP}}$ gate. These experiments gave $T_\mathrm{2} = 6.8~\mu$s as compared to $T_\mathrm{2} = 29.5~\mu$s in the absence of the drive.
We postulate that this deterioration of the dephasing time is due to thermal photons produced in the 20dB attenuator at 10mK.
These photons produce a fluctuating cavity population and dephase the qubit
as described in \cite{Rigetti12}, from which we estimated a thermal photon
number of 0.1 emitted by sources (the central pin of the bulk-head SMA connectors
used as input and output ports) at 240 mK.
Experiments with non-dissipative attenuation techniques are underway.

Finally, we perform quantum process tomography of the two qubit Clifford
operator $\mathsf{bSWAP}$ (Fig.~\ref{fig4}(c) and \ref{fig4}(d)).
The measured gate fidelity is
$F_{\mathsf{bSWAP}} = 87.3\%$ (raw) and $80\%$ (MLE).

In conclusion, we have introduced a new two-qubit gate based on the $|00\rangle \rightarrow |11\rangle$ transition, which is forbidden but can be driven by a two photon interaction.  Due to the higher levels of our system this rate can be greatly enhanced when the $\vert 0 \rangle \rightarrow \vert 1 \rangle$ transition of one transmon approaches the $\vert 1 \rangle \rightarrow \vert 2 \rangle$ transition of the other transmon. 
The resulting gate creates a maximally entangled state between two qubits directly from the ground state and, like the single qubit gates, is implemented with a single microwave pulse of defined duration, amplitude and phase. 
Together with single qubit gates this generates a universal set of gates for quantum computation.
Based on the interactions we have shown here, we believe additional two-qubit gate schemes are possible including, for example, off-resonance driving of the $|11\rangle \rightarrow |22\rangle$ transition.
Further, we have shown it is possible to realize high-fidelity quantum gates and entangled states with discrete component superconducting qubits. This general approach may hold promise as a complement to the established method of building prototype quantum processors with integrated quantum circuits.


\begin{acknowledgments}
We acknowledge fruitful discussions and contributions from David P. DiVincenzo,
Zachary Dutton, Blake R. Johnson, Colm A. Ryan, Erik Lucero and Douglas McClure
as well as Mark Ketchen for his contributions in managing the program.
We acknowledge support from IARPA under Contract No. W911NF-10-1-0324. All statements of
fact, opinion or conclusions contained herein are those of the authors and should not be
construed as representing the official views or policies of the U.S. Government.

\end{acknowledgments}


\begin{thebibliography}{10}

\bibitem{Paik11}
H.~Paik et~al., Phys. Rev. Lett. \textbf{107}, 240501 (2011).

\bibitem{Rigetti12}
C.~Rigetti et~al., arXiv:1202.5533, accepted for publication in PRB.

\bibitem{Chow12}
J.~M. Chow et~al., Phys. Rev. Lett. \textbf{109}, 060501 (2012).

\bibitem{Magesan12}
E.~Magesan et~al., Phys. Rev. Lett. \textbf{109}, 080505 (2012).

\bibitem{Martinis02}
J.~M. Martinis et~al., Phys. Rev. Lett. \textbf{89}, 117901 (2002).

\bibitem{Wallraff2005}
A.~Wallraff et~al., Phys. Rev. Lett. \textbf{95}, 060501 (2005).

\bibitem{Paraoanu06}
G.~S. Paraoanu, Phys. Rev. B \textbf{74}, 140504 (2006).

\bibitem{Rigetti10}
C.~Rigetti and M.~Devoret, Phys. Rev. B \textbf{81}, 134507 (2010).

\bibitem{Mooij10}
P.~C. de~Groot et~al., Nat. Phys. \textbf{6}, 763 (2010).

\bibitem{Chow11}
J.~M. Chow et~al., Phys. Rev. Lett. \textbf{107}, 080502 (2011).

\bibitem{Steffen2006}
M.~Steffen et~al., Science \textbf{313}, 1423 (2006).

\bibitem{Majer2007}
J.~Majer et~al., Nature (London) \textbf{449}, 443 (2007).

\bibitem{Martinis10}
R.~C. Bialczak et~al., Nat. Phys. \textbf{6}, 409 (2010).

\bibitem{Dewes12}
A.~Dewes et~al., Phys. Rev. Lett. \textbf{108}, 057002 (2012).

\bibitem{Strauch03}
F.~W. Strauch et~al., Phys. Rev. Lett. \textbf{91}, 167005 (2003).

\bibitem{DiCarlo09}
L.~DiCarlo et~al., Nature (London) \textbf{460}, 240 (2009).

\bibitem{Yamamoto2010}
T.~Yamamoto et~al., Phys. Rev. B \textbf{82}, 184515 (2010).

\bibitem{DiCarlo2010}
L.~DiCarlo et~al., Nature (London) \textbf{467}, 574 (2010).

\bibitem{Lanyon2009}
B.~P. Lanyon et~al., Nat Phys \textbf{5}, 134 (2009).

\bibitem{Bushev2010}
P. Bushev et ~al., Phys. Rev. B \textbf{82}, 134530 (2010).

\bibitem{Sorensen2000}
A.~S\o{}rensen and K.~M\o{}lmer, Phys. Rev. A \textbf{62}, 022311 (2000).

\bibitem{Benhelm2008}
J.~Benhelm et~al., Nat Phys \textbf{4}, 463 (2008).

\bibitem{Monz2011}
T.~Monz et~al., Phys. Rev. Lett. \textbf{106}, 130506 (2011).

\bibitem{Koch2007}
J.~Koch et~al., Phys. Rev. A \textbf{76}, 042319 (2007).

\bibitem{Blais07}
A.~Blais et~al., Phys. Rev. A \textbf{75}, 032329 (2007).

\bibitem{Note1}
Refs. \cite {Strauch03,DiCarlo09, Yamamoto2010,DiCarlo2010} use a similar
  condition with fast flux tuning to generate two-qubit gates.
  
\bibitem{Corcoles2011}
A.~D. Corcoles et~al., Appl. Phys. Lett. \textbf{99}, 181906  (2011).

\bibitem{Filipp09}
S.~Filipp et~al., Phys. Rev. Lett. \textbf{102}, 200402 (2009).

\bibitem{Paris04}
M.~Paris and J.~{\v{R}}eh{\'a}{\v{c}}ek, \emph{Quantum State Estimation},
  Lecture Notes in Physics (Springer, 2004).

\end{thebibliography}
\providecommand{\noopsort}[1]{}\providecommand{\singleletter}[1]{#1}%

\end{document}


\title {Supplementary material for `Entanglement of two superconducting qubits in a waveguide cavity via monochromatic two-photon excitation'}
\author{S. Poletto}
\author{Jay M. Gambetta}
\author{Seth T. Merkel}
\author{John A. Smolin}
\author{Jerry M. Chow}
\author{A.D. C\'orcoles}
\author{George A. Keefe}
\author{Mary B. Rothwell}
\author{J.R. Rozen}
\author{D.W. Abraham}
\author{Chad Rigetti}
\author{M. Steffen}
\affiliation{IBM T.J. Watson Research Center, Yorktown Heights, NY 10598, USA}

\date{\today}

\pacs{03.67.Ac, 03.67.Bg, 42.50.Pq, 85.25.-j}

\maketitle

\section{Outline of  the Schrieffer-Wolff transformation}

In this section we outline the Schrieffer-Wolff transformation \cite{schrieffer}
used to derive the effective Hamiltonian. We start by making the assumption that
the general Hamiltonian is of the form $H=H_0+\lambda V$ where $\lambda$ is a
small parameter and $H_0$ is some free Hamiltonian.
An effective Hamiltonian for the system is $H_\mathrm{eff} = A^\dagger H A$ where $A = \exp(-i S)$.
Assuming a power series expansion of $S$ (which we restrict to be at least first order in $\lambda$)    
\begin{equation} \label{frameS}
S = \sum\limits_{n=1}^{\infty} S^{(n)}\lambda^n,
\end{equation} 
we derive $H_\mathrm{eff}$ from the Baker-Campbell-Hausdorff formula as 
\begin{equation}
\begin{split}
H_\mathrm{eff} =& \sum\limits_{j=0}^{\infty} \mathrm{ad}[i\sum\limits_{n=1}^{\infty} S^{(n)}\lambda^n]^j\frac{H_0+\lambda V}{j!},\\=
&\sum\limits_{m=0}^{\infty}\lambda^{m} H^{(m)},
\end{split}
\end{equation}
where $H^{(0)}=H_0$ and $\mathrm{ad}(\cdot)$ denotes the adjoint representation $\mathrm{ad}(A)B = [A,B]$.
For $m>0$ 
\begin{equation}\label{relation}
H^{(m)} = i[S^{(m)},H_0]+H_\mathrm{x}^{(m)},
\end{equation} with
\begin{equation}
\begin{split}
H_\mathrm{x}^{(1)} =& V,\\
H_\mathrm{x}^{(2)} =& -[S^{(1)},[S^{(1)},H_0]]/2+i[S^{(1)},V],\\
H_\mathrm{x}^{(3)} =& -[S^{(1)},[S^{(2)},H_0]]/2-[S^{(2)},[S^{(1)},H_0]]/2
-i[S^{(1)},[S^{(1)},[S^{(1)},H_0]]]/3!+i[S^{(2)},V]-[S^{(1)},[S^{(1)},V]]/2.
\end{split}
\end{equation}
Since $H_{\rm x}^{(m)}$ depends on $S^{(j)}$ only up to $j = m-1$,
we can solve \eqrf{relation} for $S^{(m)}$ at each order by enforcing
that the effective Hamiltonian $H^{(m)}$ describes our desired dynamics.
 
For example if we are interested in diagonalizing the system ($H^{(m)}$ to be diagonal in the basis of $H_0$) then we find
\begin{equation}\label{relation2}
\sum_{p}E^{(m)}_p \ket{p}\bra{p} = i\sum_{p}E^{(0)}_p(S^{(m)} \ket{p}\bra{p}-\ket{p}\bra{p}S^{(m)})+ H_\mathrm{x}^{(m)},
\end{equation}
where $E^{(m)}_p  = \bra{p}H_\mathrm{x}^{(m)}\ket{p}$ are the eigenvalues of $H_\mathrm{x}^{(m)}$,
which gives
\begin{equation}\label{relation3}
\bra{p}S^{(m)} \ket{q}  = \frac{-i\bra{p}H_\mathrm{x}^{(m)}\ket{q}}{E^{(0)}_p-E^{(0)}_q}.
\end{equation}


Another example of interest is when we want to find an effective Hamiltonian after we have eliminated a high energy subspace, that is we want $H^{(m)} =H^{(m)}_l\oplus H^{(m)}_h$ where $l$ labels the low energy subspace and $h$ the high energy subspace.   In this case Eq. \eqref{relation} becomes
\begin{equation}
\begin{split}
\begin{pmatrix}
H^{(m)}_l & 0 \\
0 & H^{(m)}_h
\end{pmatrix} &= i \begin{pmatrix}
S^{(m)}_{ll} & S^{(m)}_{lh} \\
S^{(m)}_{hl} &S^{(m)}_{hh}
\end{pmatrix}\begin{pmatrix}
H_{0_l} & 0 \\
0 & H_{0_h}
\end{pmatrix}
-i\begin{pmatrix}
H_{0_l} & 0 \\
0 & H_{0_h}
\end{pmatrix}\begin{pmatrix}
S^{(m)}_{ll} & S^{(m)}_{lh} \\
S^{(m)}_{hl} &S^{(m)}_{hh}
\end{pmatrix} +\begin{pmatrix}
H_\mathrm{x_{ll}}^{(m)} &H_\mathrm{x_{lh}}^{(m)} \\
H_\mathrm{x_{hl}}^{(m)} &H_\mathrm{x_{hh}}^{(m)}
\end{pmatrix}.
\end{split}
\end{equation}
Here we have assumed $H_0 =H_{0_l}\oplus H_{0_h}$. This can be expanded to give
\begin{equation}
\begin{split}
H^{(m)}_l=&iS^{(m)}_{ll}H_{0_l}-iH_{0_l}S^{(m)}_{ll}
+H_\mathrm{x_{ll}}^{(m)},  \\
0=&i S^{(m)}_{lh}H_{0_h}-iH_{0_l}S^{(m)}_{lh}+H_\mathrm{x_{lh}}^{(m)},\\
0=&iS^{(m)}_{hl}H_{0_l}-iH_{0_h}S^{(m)}_{hl}+H_\mathrm{x_{hl}}^{(m)},\\
H^{(m)}_h=&iS^{(m)}_{hh}H_{0_h}-iH_{0_h}S^{(m)}_{hh}+H_\mathrm{x_{hh}}^{(m)},
\end{split}
\end{equation}
and without loss of generality we can take $iS^{(m)}_{ll}=iS^{(m)}_{hh}=0$ so that
\begin{equation}
\begin{split}
H^{(m)}_l=&H_\mathrm{x_{ll}}^{(m)},  \\
H_{0_l}S^{(m)}_{lh}-S^{(m)}_{lh}H_{0_h}=&-iH_\mathrm{x_{lh}}^{(m)},\\
H_{0_h}S^{(m)}_{hl}-S^{(m)}_{hl}H_{0_l}=&-iH_\mathrm{x_{hl}}^{(m)},\\
H^{(m)}_h=&H_\mathrm{x_{hh}}^{(m)},
\end{split}
\end{equation} and if $H_{0}$ is diagonal then 
\begin{equation} \label{highlow}
\begin{split}
\bra{p} S^{(m)}_{lh}\ket{q}=&\frac{-i \bra{p} H_\mathrm{x_{lh}}^{(m)}\ket{q}}{\bra{p} H_{0_l}\ket{p}-\bra{q}H_{0_h}\ket{q}},\\
\bra{p} S^{(m)}_{hl}\ket{q}=&\frac{-i\bra{p} H_\mathrm{x_{hl}}^{(m)}\ket{q}}{\bra{p} H_{0_h}\ket{p}-\bra{q}H_{0_l}\ket{q}}.
\end{split}
\end{equation}

\section{Derivation of the two-photon gate}

We describe the free evolution of the system by the Hamiltonian 
\begin{equation}
\begin{split}
H_\mathrm{sys} =& (\omega_1 - \frac{\delta_1}{2}) a^\dagger a + \frac{\delta_1}{2} (a^\dagger a)^2 +(\omega_2 - \frac{\delta_2}{2}) b^\dagger b + \frac{\delta_2}{2} (b^\dagger b)^2+J(a^\dagger b + a b^\dagger),
\end{split}
\end{equation} where  $\omega_{1(2)}$ is the $|0\rangle \rightarrow |1\rangle$
transition frequency of transmon 1(2), $\delta_{1(2)}$ is the anharmonicity of
transmon 1(2),  $J$ is the effective strength of the exchange interaction between
transmons, and $a(b)$ are the annihilation operators for transmon 1(2).
To represent external driving of the system we use the Hamiltonian 
\begin{equation}
H_\mathrm{c}= [\Omega_1(a+a^\dagger)+\Omega_2(b+b^\dagger)]\cos(\omega_d t + \phi).\label{eq:hamiltonian}
\end{equation}
where $\Omega_{1(2)}$ is the amplitude of the applied microwave field of frequency
$\omega_d$ and phase $\phi$ to transmon 1(2). 

To derive the effective Hamiltonian for the two-photon transition $|00\rangle \rightarrow |11\rangle$
we perform the following procedure. First we use Schrieffer-Wolff transformation to go to a frame which diagonalizes the system Hamiltonian $H_\mathrm{sys}$ to second order in $J/\Delta$ where $\Delta = \omega_1-\omega_2$. Using \eqrf{relation3} we find the eigenenergies and the $S$ operator.  This then gives the frame transformation $A$ and the transferred controls are found using $A^\dagger H_\mathrm{c}A$. The next step is to move to a frame at the drive frequency which is close to $\omega_d = (\omega_1+\omega_2)/2 - \delta$ and then finally we separate the levels into high and low energy manifolds. In the case when $\Delta$ is approximately equal to $\delta_2$ the low subspace is spanned by the states $\ket{00}$, $\ket{11}$, and $\ket{02}$ (it would switch to $\ket{20}$ if $\Delta$ approaches $-\delta_1$). Here we can use  another Schrieffer-Wolff transformation to eliminate the coupling between theses manifolds. This is done using \eqrf{highlow}.
This whole procedure is performed using \textit{Mathematica}$^\textrm{\textregistered}$ \cite{mathematica}
and returns a second order effective Hamiltonian in the two-qubit subspace of the form
\begin{equation} \label{eff}
\begin{split}
H_\mathrm{eff} = &\frac{\alpha_{IZ}}{2} IZ +\frac{\alpha_{ZI}}{2} ZI+\frac{\alpha_{ZZ}}{4} ZZ+\frac{\Omega_{BR}}{4} (XX-YY)
+\frac{\Omega_{BI}}{4} (XY+YX) +\frac{\Omega_{S}}{2} (XX+YY),
\end{split}
\end{equation} 
where 
\begin{equation}
\begin{split}
\Omega_{BR}=&\cos(2\phi)\Omega_B,
\\
\Omega_{BI}=&\sin(2\phi)\Omega_B,
\\
\Omega_{S}=&\frac{-2 J \delta  \left(J \Omega_1 \Omega_2 (\delta_1-\delta_2)+\Omega_2^2 \delta_2 (\delta_1+\Delta)+\Omega_1^2 \delta_1 (-\delta_2+\Delta)\right)}{\Delta(\delta_2-\Delta) (\delta_1+\Delta)  \left(-4 \delta ^2+\Delta^2\right)},
\end{split}
\end{equation} 
with
\begin{equation}
\Omega_{B} = \frac{-2 J \left(-J \Omega_1 \Omega_2 (\delta_1+\delta_2)+\Omega_2^2 \delta_2 (\delta_1+\Delta)+\Omega_1^2 \delta_1 (\delta_2-\Delta)\right)}{(\delta_2-\Delta) (\delta_1+\Delta) \left(-4 \delta ^2+\Delta^2\right)}.
\end{equation} 
This last equation, in the limit $\delta \ll \Delta,$ reduces to the one used in the main text.
We find rather long expressions for $\alpha_{IZ}$, $\alpha_{ZI}$ and  $\alpha_{ZZ}$. They are approximately (to first order in $J$) given by 
\begin{equation}
\begin{split}
\alpha_\mathrm{ZZ} = &-\frac{\Omega_2^2}{2 (\delta +\delta_2)-\Delta}-\frac{2 \left(\left(-8 \delta ^2 (\delta +\delta_1)+16 \delta_1 \delta_2^2-4 \left(\delta ^2+4 \delta_1 \delta_2\right) \Delta+2 (\delta +\delta_1) \Delta^2+\Delta^3\right) \Omega_1 \Omega_2\right) J}{(\delta_2-\Delta) (2 (\delta +\delta_2)-\Delta) (2 (\delta +\delta_1)+\Delta) \left(-4 \delta ^2+\Delta^2\right)}\\
\frac{\alpha_\mathrm{IZ}+\alpha_\mathrm{ZI}}{2} = &\left(-\delta -\frac{\delta_1 \Omega_1^2}{(2 \delta +\Delta) (2 (\delta +\delta_1)+\Delta)}-\frac{\delta_2 \Omega_2^2}{4 \delta  (\delta +\delta_2)-2 (2 \delta +\delta_2) \Delta+\Delta^2}\right)\\
&+\frac{4 (2 \delta  (\delta_1+\delta_2)+(-\delta_1+\delta_2) \Delta) \Omega_1 \Omega_2 J}{(2 (\delta +\delta_2)-\Delta) (2 (\delta +\delta_1)+\Delta) \left(4 \delta ^2-\Delta^2\right)}\\
\frac{\alpha_\mathrm{IZ}-\alpha_\mathrm{ZI}}{2} = &\frac{1}{2} \left(\Delta+\frac{2 \delta_1 \Omega_1^2}{(2 \delta +\Delta) (2 (\delta +\delta_1)+\Delta)}-\frac{\Omega_2^2}{2 \delta -\Delta}\right)\\
&+\frac{\left(4 \delta ^2+4 \delta  (\delta_1-\delta_2+\Delta)+\Delta (-2 (\delta_1+\delta_2)+\Delta)\right) \Omega_1 \Omega_2 J}{(\delta_2-\Delta) (2 (\delta +\delta_1)+\Delta) \left(-4 \delta ^2+\Delta^2\right)}.
\end{split}
\end{equation}

We note that the three operators $(XX-YY)/2$, $(XY+YX)/2$ and $(ZI+IZ)/2$
form a SU(2) algebra and that both $ZZ$ and $IZ-ZI$ commute with all three
of these operators.  Also, $\Omega_S$ is smaller than $\Omega_B$ by a factor of
$\delta/\Delta$ which leaves only the term described by $\alpha_{IZ}+\alpha_{ZI}$
in Eq.~\eqref{eff} as a non-trivial error.
Therefore, we choose $\delta$ such that $\alpha_{IZ}+\alpha_{ZI} =0$ which corresponds,
for a given power, to the drive frequency that shows the largest amplitude oscillations
in two-photon spectroscopy. This leads to the unitary evolution derived from
the effective Hamiltonian as
\begin{equation}
U = U_{B} U_{ZZ}U_{IZ-ZI}
\end{equation} where
\begin{equation}
U_{B}=\begin{pmatrix}
\cos(\frac{\Omega_\mathrm{B} t}{2})  & 0 & 0 & -i e^{-2 i \phi} \sin(\frac{\Omega_\mathrm{B} t}{2}) \\
0 & 1 & 0 & 0 \\
0 & 0 & 1 & 0 \\
-i e^{2 i \phi}\sin(\frac{\Omega_\mathrm{B} t}{2})  & 0 & 0 &  \cos(\frac{\Omega_\mathrm{B} t}{2}) \\
\end{pmatrix}
\end{equation}
$U_{ZZ}= \exp(-i \alpha_{ZZ} ZZ t/4)$ and $U_{IZ-ZI}=\exp(-i ( \alpha_{IZ}- \alpha_{ZI}) (IZ-ZI) t/4)$.


\providecommand{\noopsort}[1]{}\providecommand{\singleletter}[1]{#1}%